\documentclass[10pt]{article}
\usepackage{graphicx}

\def\Title#1{\begin{center} {\Large #1 } \end{center}}
\def\Author#1{\begin{center}{ \sc #1} \end{center}}
\def\Address#1{\begin{center}{ \it #1} \end{center}}

\newcommand\pubblock{\rightline{\begin{tabular}{l} Proceedings of the Fifth Annual LHCP\\ \pubnumber\\
         \pubdate  \end{tabular}}}

\newenvironment{Abstract}{\begin{quotation} \begin{center} 
             \large ABSTRACT \end{center}\bigskip 
      \begin{center}\begin{large}}{\end{large}\end{center} \end{quotation}}

\newenvironment{Presented}{\begin{quotation} \begin{center} 
             PRESENTED AT\end{center}\bigskip 
      \begin{center}\begin{large}}{\end{large}\end{center} \end{quotation}}

\def\Acknowledgements{\bigskip  \bigskip \begin{center} \begin{large}
             \bf ACKNOWLEDGEMENTS \end{large}\end{center}}

\def\beq{\begin{equation}}
\def\eeq#1{\label{#1}\end{equation}}
\def\eeqn{\end{equation}}

\def\beqa{\begin{eqnarray}}
\def\eeqa#1{\label{#1}\end{eqnarray}}
\def\eeqan{\end{eqnarray}}

\let\bar=\overbar

\def\Dslash{\not{\hbox{\kern-4pt $D$}}}
\def\dslash{\not{\hbox{\kern-2pt $\del$}}}

\def\msb{{\bar{\ssstyle M \kern -1pt S}}}

\usepackage{color}
\usepackage[T1]{fontenc}
\usepackage{ptdr-definitions}
\usepackage{url}

\newcommand\pubnumber{ CMS-CR-2017/368 }

\newcommand\pubdate{\today}

\def\affiliation{
On behalf of the ATLAS and CMS Collaborations, \\
Center for Scientific Computing \\
S\~ao Paulo State University, S\~ao Paulo, Brazil}

\def\supporttext{This material is based upon work supported by the S\~ao Paulo Research Foundation (FAPESP) under grants No.~2013/01907-0 and 2016/15897-4.}
\def\support{\footnote{\support}}

\let\OLDthebibliography\thebibliography
\renewcommand\thebibliography[1]{
  \OLDthebibliography{#1}
  \setlength{\parskip}{2.0pt plus 0.5ex}
  \setlength{\itemsep}{2.0pt plus 0.5ex}
}

\begin{document}

\large
\begin{titlepage}
\pubblock

\vfill
\Title{  Evolution of online algorithms in ATLAS and CMS in Run~2  }
\vfill

\Author{ Thiago R. F. P. Tomei }
\Address{\affiliation}
\vfill
\begin{Abstract}

The Large Hadron Collider has entered a new era in Run~2, with centre-of-mass energy of 13~TeV and instantaneous luminosity reaching $\mathcal{L}_\textrm{inst} = 1.4\times$10\textsuperscript{34}~cm\textsuperscript{-2}~s\textsuperscript{-1} for pp collisions. In order to cope with those harsher conditions, the ATLAS and CMS collaborations have improved their online selection infrastructure to keep a high efficiency for important physics processes -- like W, Z and Higgs bosons in their leptonic and diphoton modes -- whilst keeping the size of data stream compatible with the bandwidth and disk resources available. In this note, we describe some of the trigger improvements implemented for Run~2, including algorithms for selection of electrons, photons, muons and hadronic final states.
\end{Abstract}
\vfill

\begin{Presented}
The Fifth Annual Conference\\
 on Large Hadron Collider Physics \\
Shanghai Jiao Tong University, Shanghai, China\\ 
May 15-20, 2017
\end{Presented}
\vfill
\end{titlepage}
\def\thefootnote{\fnsymbol{footnote}}
\setcounter{footnote}{0}
%

\normalsize 


\section{Introduction}

It is widely known that, for high-luminosity pp operations at the LHC, it is neither possible nor desirable to record every single collision, both due to bandwidth and data storage constraints. Both the ATLAS~\cite{ATLASPaper} and CMS~\cite{CMSPaper} experiments employ \emph{trigger systems} that do a quasi-realtime (online) analysis of the collected data stream to select the most interesting events for permanent recording. For Run~2, both experiments employ a two-tiered system: a Level-1 trigger, implemented in hardware with partial detector readout, that reduces the data rate to $\mathcal{O}$(100~kHz); and a High-Level Trigger, implemented in software with full detector readout, that further reduces the data rate to $\mathcal{O}$(1~kHz). The bandwidth allocation is generally driven by physics priorities in both experiments, with most resources allocated to inclusive triggers that can be used for many different studies in the experiment. The complexity of the online selection is comparable to that of the offline analysis: in ATLAS, 2000 active trigger chains were running in 2016 pp data taking, whilst CMS had $\mathcal{O}$(300) L1 seeds and $\mathcal{O}$(500) HLT paths for the same period.

\section{ATLAS and CMS Trigger Improvements for Run~2}

\begin{figure}[htbp]
   \centering
   \includegraphics[height=2in]{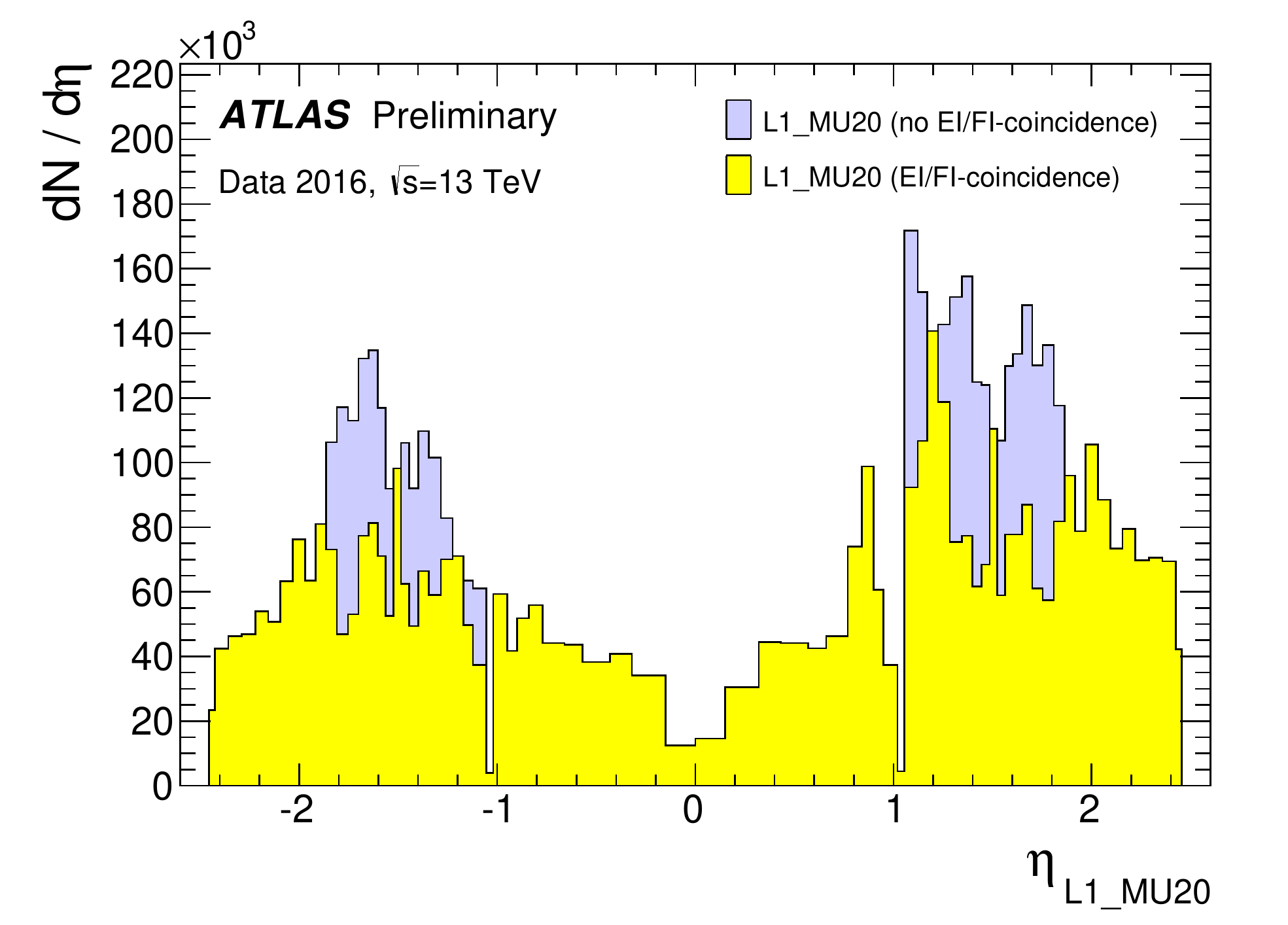}
   \includegraphics[height=2in]{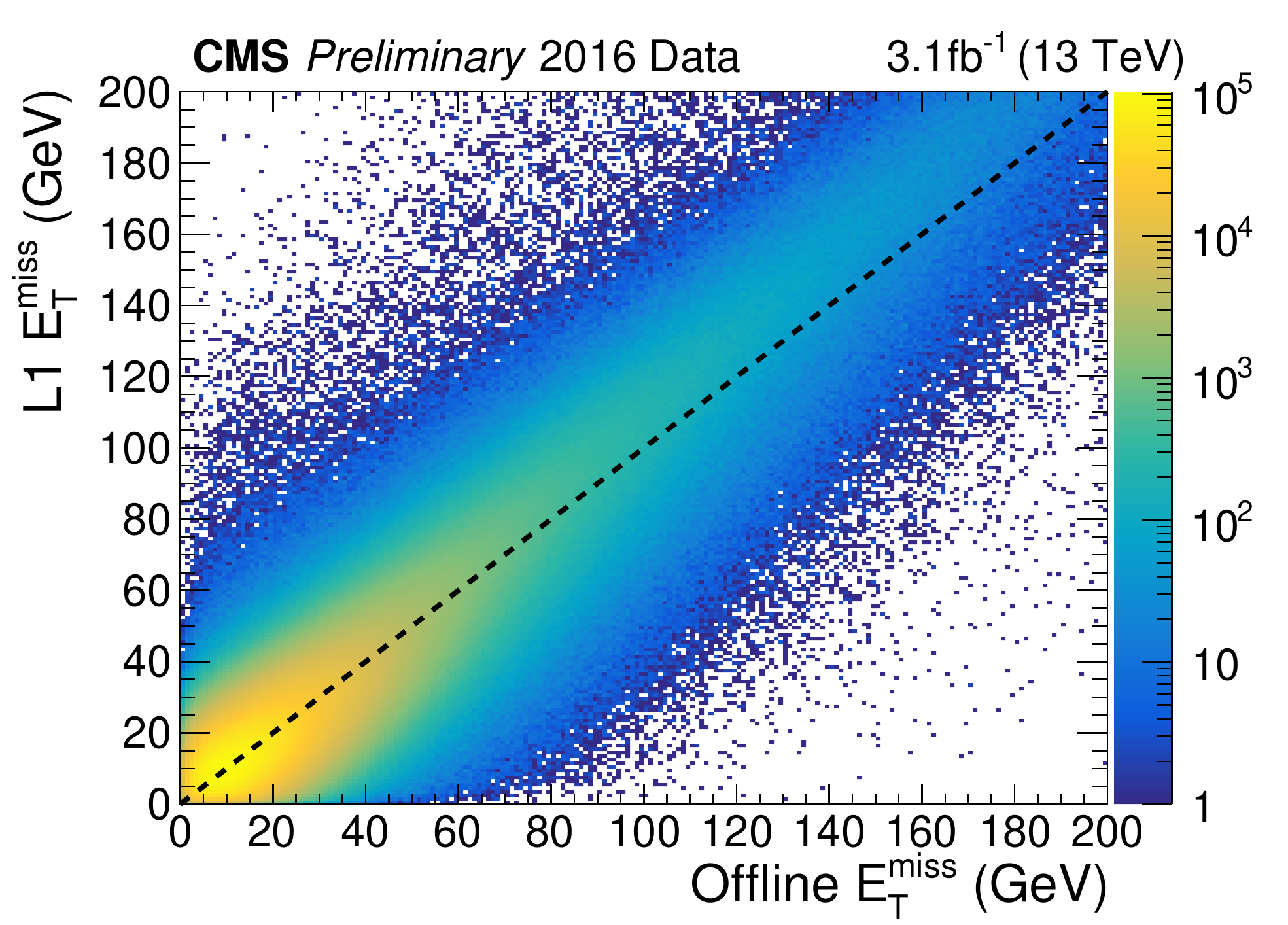} 
   \caption{ATLAS and CMS L1 trigger improvements. Left: rate reduction in ATLAS endcap muon triggers thanks to additional coincidence with TGC inner chambers~\cite{ATLASTrigger:2016}. Right: correlation between \MET{} calculated at L1 and in the offline analysis, demonstrating the effectiveness of the event-by-event pileup subtraction~\cite{CMSL1Calo:2016}.}
   \label{fig:L1Improvements}
\end{figure}

For Run~2, both experiments have upgraded their trigger systems to cope with the increased centre-of-mass energy and luminosity conditions. ATLAS has deployed an improved topological trigger (L1Topo) that allows finer selection on quantities from L1Calo and L1Muon as well as on composite quantities like \MET{} and \HT{}. Both the L1Calo and the L1Muon systems have been improved as well; L1Calo now is equipped with digital autocorrelation Finite Impulse Response (FIR) filters, the capability to do dynamic, bunch-by-bunch pedestal correction and a new set of energy-dependent L1 EM isolation criteria for the cluster processor. The L1Muon system, on the other hand, has been improved to require additional coincidence with TGC inner chambers to reduce trigger rates in the endcap (see Fig.~\ref{fig:L1Improvements}), whilst new ROC chambers enhanced the trigger coverage by 4\% in the barrel region. ATLAS has also merged the L2 and Event Filter farms to allow for more flexible optimisations, moving away from the three-tiered system they had during Run~1.

CMS has also improved their systems: their Level-1 Calorimeter Trigger now does event-by-event pileup subtraction (see Fig.~\ref{fig:L1Improvements}) and is equipped with advanced algorithms for dynamic clustering; for electron/photon selection, that allows for bremsstrahlung recovery, whilst for tau selection it allows to merge the treatment of multiprong decays and isolation. On the Level-1 Muon side, CMS now combines the subdetectors (DT, RPC, CSC) information at an earlier stage, optimising the selection separately for three regions of barrel, overlap and endcap. On the High-Level Trigger front, CMS migrated their software framework to a full multithreaded model in 2016, and have completely reoptimised their track reconstruction for the Phase-1 pixel upgrade that happened in 2017.

\clearpage

\section{Electron and Photon Triggers}

Triggering on events containing electrons and photons in the LHC collisions is complicated by the large backgrounds from multijet events; a hadronic jet can be easily misidentified as a single e/$\gamma$, especially if enriched in electromagnetic component. To mitigate that phenomenon, the experiments deploy identification and isolation algorithms already at the online selection, both at the L1 trigger and at the HLT. The prototype trigger algorithm is the \emph{single electron trigger}, which generally tries to reconstruct and identify the energy deposit in the electromagnetic calorimeter and match it with a charged particle track; photon triggers are similarly built, but without the tracking steps.

\begin{figure}[htb]
\centering
\includegraphics[height=1.9in]{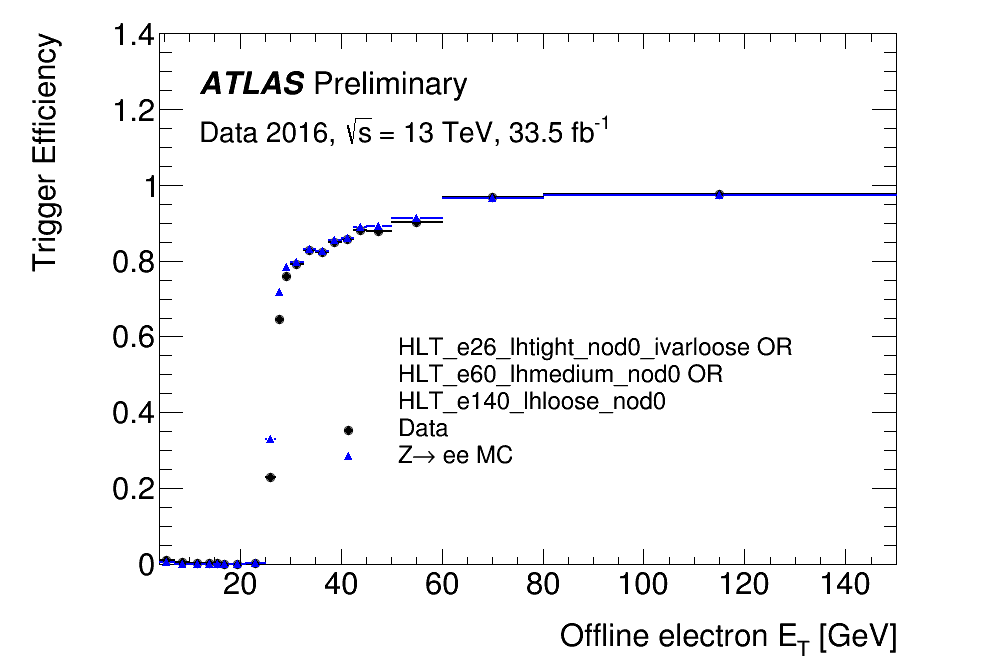}
\includegraphics[height=1.9in]{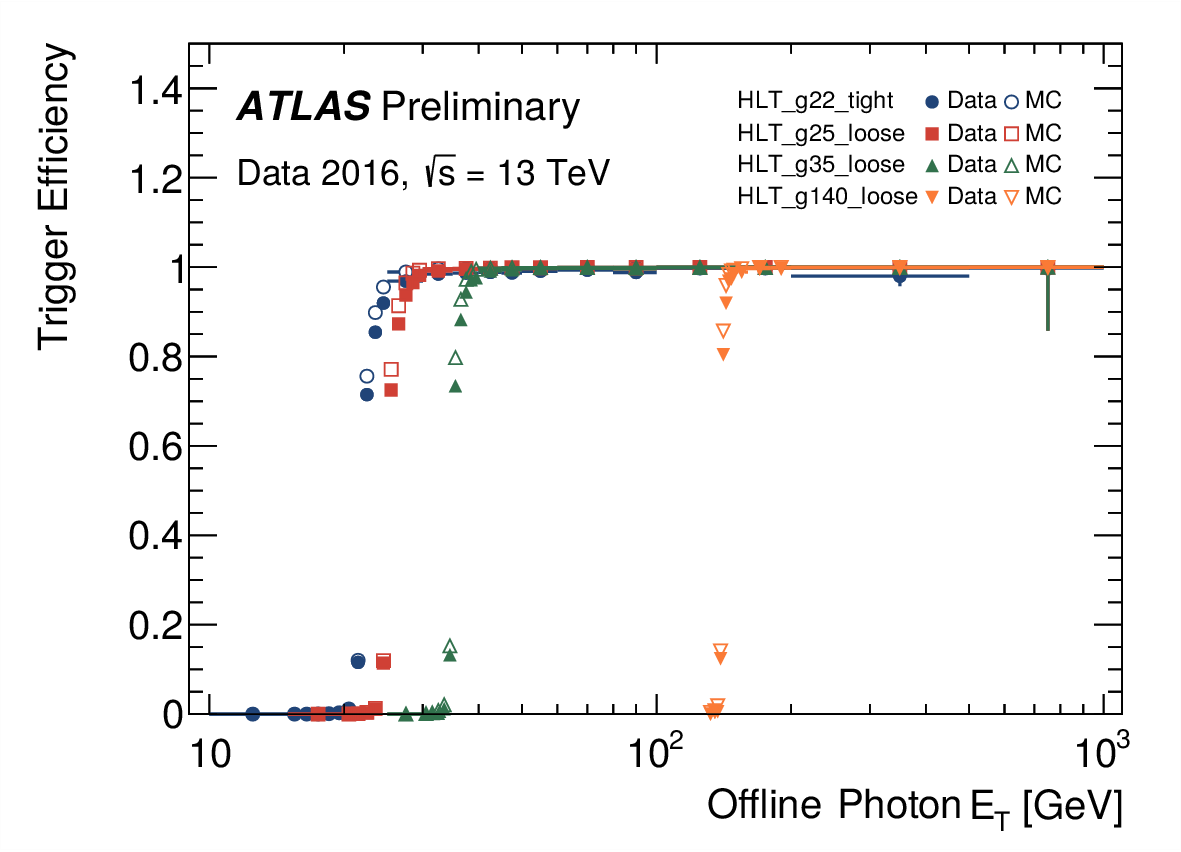}
\caption{Left: ATLAS turn-on curve for the inclusive OR of the three electron triggers:
\texttt{HLT\_e26\_\allowbreak{}lhtight\_\allowbreak{}nod0\_ivarloose},
\texttt{HLT\_e60\_lhmedium\_nod0},
\texttt{HLT\_e140\_lhloose\_nod0}~\cite{ATLASOverallTRG:2016}.
Right: ATLAS turn-on curves for single photon triggers and legs of diphoton triggers:
\texttt{HLT\_g22\_tight},
\texttt{HLT\_g25\_tight},
\texttt{HLT\_g35\_tight},
\texttt{HLT\_g140\_tight}~\cite{ATLASOverallTRG:2016}.}
\label{fig:ATLASElectronPhotonTurnOn}
\end{figure}

In order to cope with the harsher conditions, ATLAS moved from a cut-based identification procedure in Run~1 to a more sophisticated, likelihood-based identification in Run~2. For isolation, at L1 they rely on the energy in the EM calorimeter deposited in a ring around the electron cluster, whilst at HLT they calculate isolation based on tracks located within a variable-size cone around the reconstructed e/$\gamma$. Three trigger algorithms were primarily used by ATLAS for electrons in 2016: a low-\PT{} trigger (26~GeV) with tight identification and isolation criteria; a medium-\PT{} trigger (60~GeV) with medium identification; and a high-\PT{} trigger (140~GeV) with loose identification. \emph{Photon triggers} follow generally the same strategy, with low-\PT{} single legs of diphoton triggers (22, 25 and 35~GeV) and a high-\PT{} (140~GeV) single photon trigger. Example performance plots for ATLAS e/$\gamma$ triggers can be seen in Fig.~\ref{fig:ATLASElectronPhotonTurnOn}.

\begin{figure}[htb]
\centering
\includegraphics[height=1.9in]{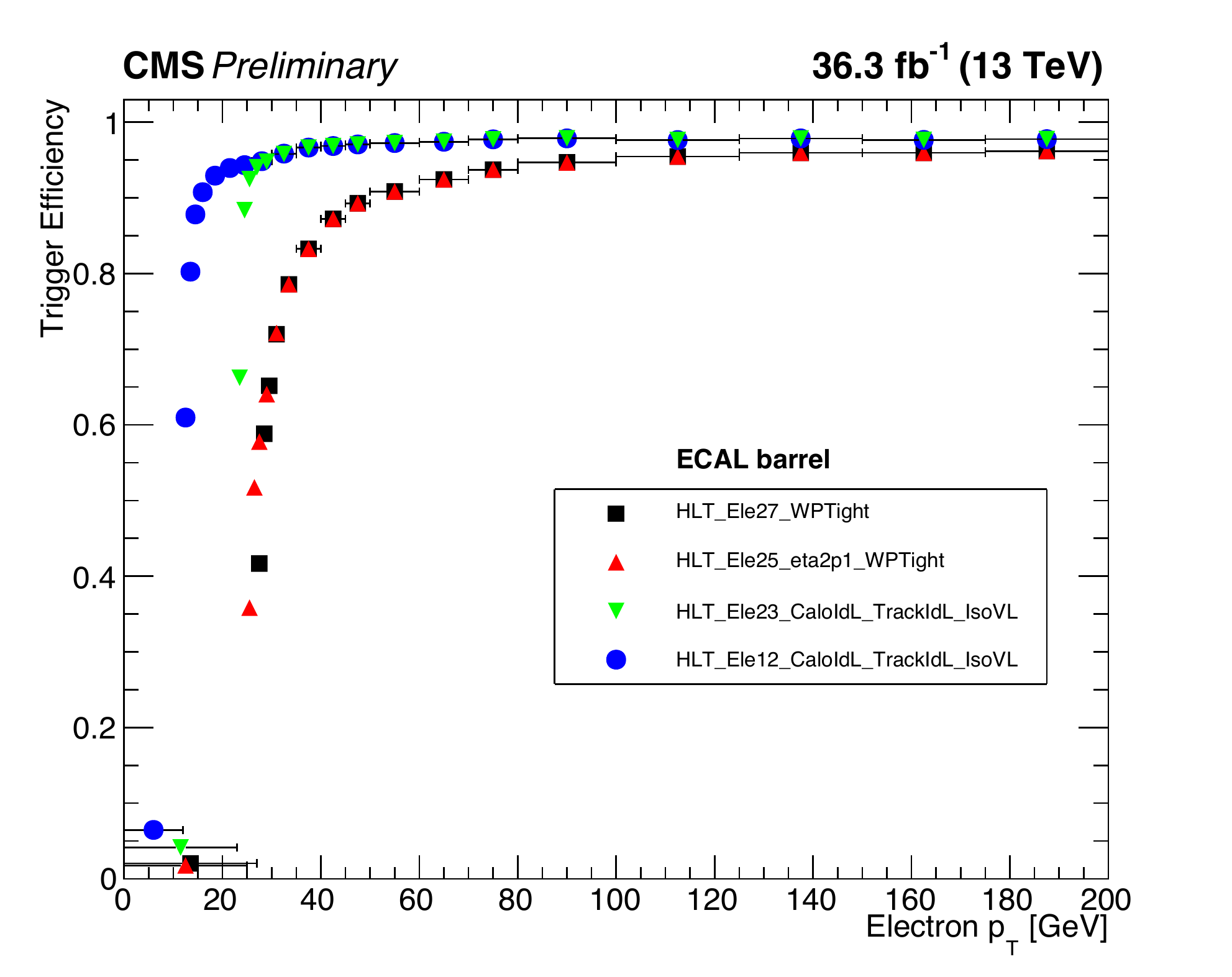}
\includegraphics[height=1.9in]{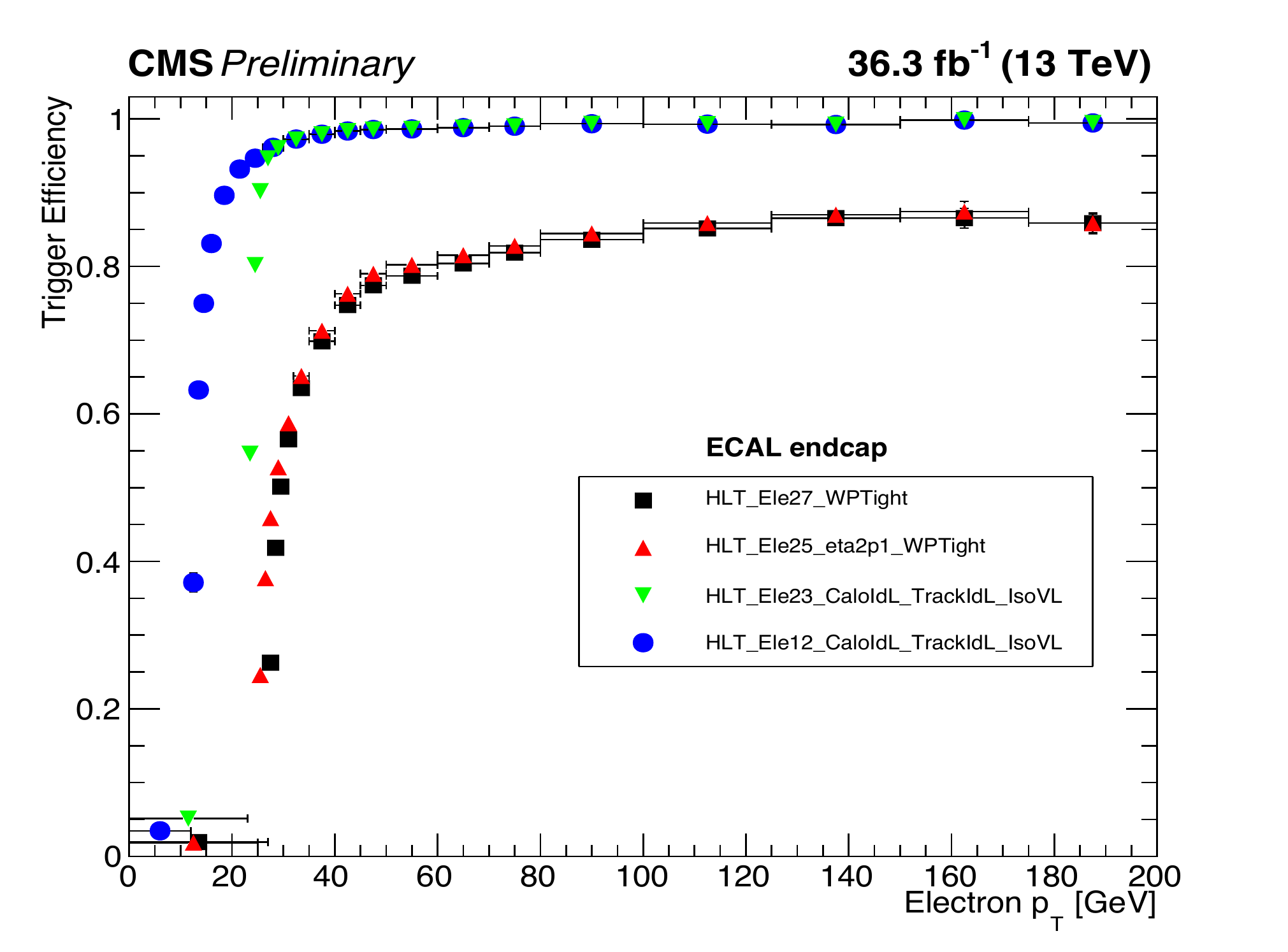}
\caption{CMS turn-on curves for the single electron trigger paths
\texttt{HLT\_Ele27\_WPTight},
\texttt{HLT\_Ele25\_eta2p1\_WPTight}
and for the single legs of the double electron trigger paths
\texttt{HLT\_Ele23\_CaloIdL\_TrackIdL\_IsoVL},
\texttt{HLT\_Ele12\_CaloIdL\_TrackIdL\_IsoVL}, for the electromagnetic calorimeter barrel (left) and endcap (right)~\cite{CMSHLTEGamma:2016}.}
\label{fig:CMSElectronTurnOn}
\end{figure}

CMS deploys a similar strategy for e/$\gamma$ triggers. In 2016 three main single-electron trigger paths were used. Two of those paths, with low \PT{} thresholds (27 and 25~GeV), had tight identification and isolation criteria based both on calorimeter and silicon tracker information, with the former path having full tracker coverage and the latter trading restricted coverage in pseudorapidity ($|\eta| < 2.1$) for a lower threshold in \PT{}. The third path is primarily aimed at searches for new physics and eschews isolation completely, but has to compensate with a very high transverse momentum threshold (105~GeV). Triggers for double electrons, single and double photons generally follow the same lines; CMS also deploys both low-\PT{}, double photon paths with selections on the invariant mass and high-\PT{}, single photon paths. Example performance plots for CMS electron triggers can be seen in Fig.~\ref{fig:CMSElectronTurnOn}.

\section{Muon Triggers}

Both CMS and ATLAS deploy trigger paths that select events containing muons, specialized for different kinds of physics: prompt muons for electroweak, top and Higgs studies, low energy muons for B physics, amongst others. The CMS HLT system considers two types of reconstructed muons: for \emph{standard muons} the system reconstructs hits in the muon system and propagates them back to the silicon tracker. For \emph{tracker muons}, instead, the HLT runs a regional iterative tracking algorithm, matched to muon system chambers. The ATLAS HLT also has two reconstruction strategies: ``muon system + inner detector'' combined as the standard reconstruction and muon system \emph{stand-alone muons} for special uses. In addition to local area reconstruction seeded from L1 location, ATLAS also deploys multi-muon triggers that, upon firing of the single L1 muon trigger, search the full detector for additional muons -- the \emph{full-scan algorithm}.  

For the benchmark single muon trigger, CMS adopts a two-pronged strategy. For triggering muons with low \PT{}, both for standard model physics and most new physics searches, the experiment deploys paths with isolation both for the standard and tracker muon reconstruction. CMS also triggers on high \PT{} muons with no isolation for special cases like boosted Z bosons and lepton+jets searches. For Run~2 triggers, CMS changed their isolation strategy to have independent selections on track, electromagnetic and hadronic isolations instead of a combined one, leading to an increase in efficiency. 

\begin{figure}[htb]
\centering
\includegraphics[height=1.9in]{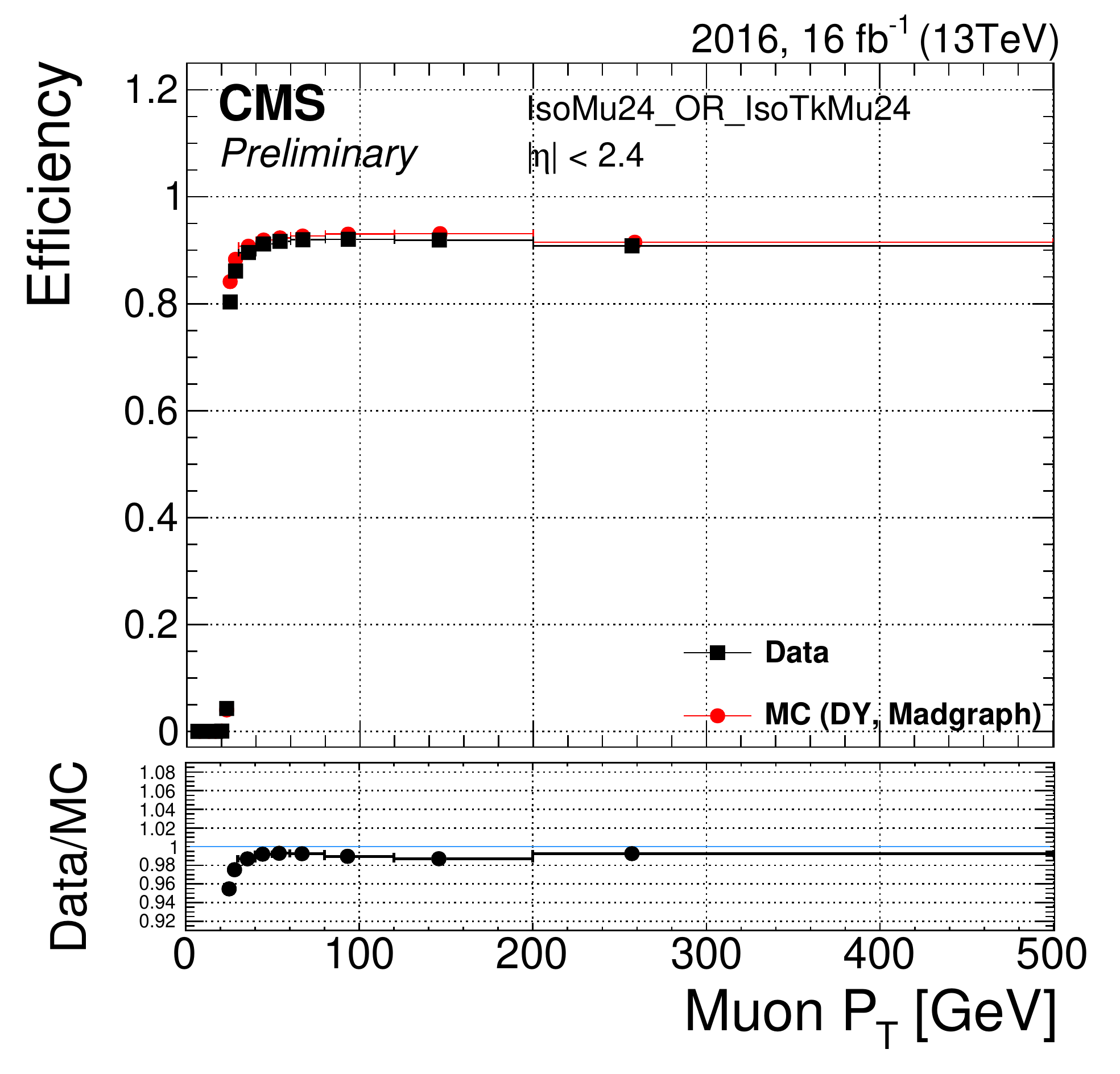}
\includegraphics[height=1.9in]{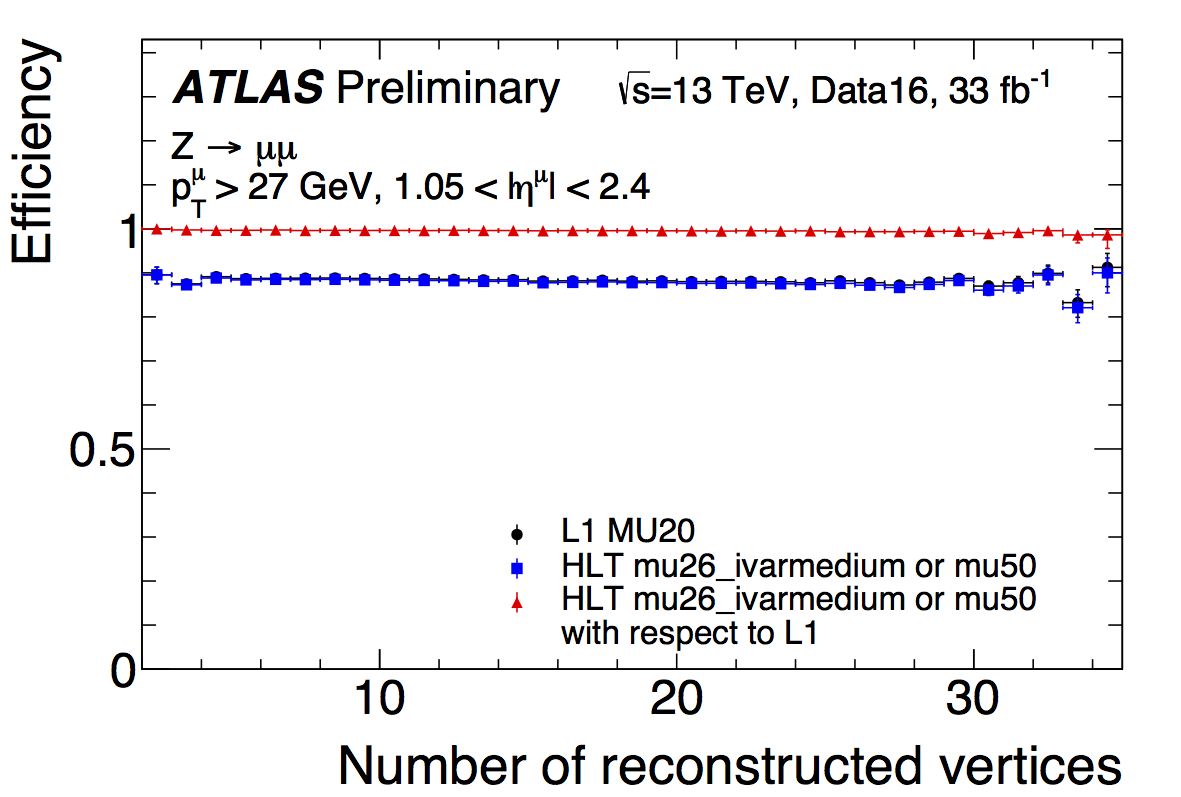}
\caption{Left: CMS turn-on curve for the inclusive OR of the \texttt{HLT\_IsoMu24} and \texttt{HLT\_IsoTkMu24} paths that implement the two kinds of reconstruction strategy (standard and tracker muons) adopted by CMS; both paths are seeded by a L1 muon with $\PT{} > 22 \GeV$~\cite{CMSHLTRun2:2016}. Right: ATLAS trigger efficiency as function of the number of reconstructed vertices for the L1 seed \texttt{L1\_MU20} and for the \texttt{HLT\_mu26\_ivarmedium} OR \texttt{HLT\_mu50} algorithms in the $1.05 < |\eta^\mu| < 2.4$ region, demonstrating the robustness against pileup effects~\cite{ATLASOverallTRG:2016}.}
\label{fig:MuonTurnOn}
\end{figure}

ATLAS adopts a very similar strategy, having in 2016 a trigger selection \texttt{mu26\_imedium OR mu50} seeded by L1 muons with $\PT > 20\GeV$. They deploy muon isolation based on inner detector tracks and for 2016 they innovate it with \emph{variable cone isolation}, with the radius depending on the muon \PT{}, whilst in 2015 and Run-1 they used fixed cone sizes; this improvement lead to a more robust performance against pileup. Example performance plots for muon triggers for both experiments can be seen in Fig.~\ref{fig:MuonTurnOn}.

\clearpage

\section{Hadronic Triggers}

Both ATLAS and CMS have dedicated trigger paths to select collisions with high energy hadronic jets. The prototype trigger algorithm is the \emph{single jet trigger}. Both experiments reconstruct jets with the anti-\kt{} algorithm with different radii for jets from standard QCD production ($R$ = 0.4) and for hadronic decays of boosted massive objects that are reconstructed as a single jet ($R$ = 1.0 for ATLAS, 0.8 for CMS). 

The ATLAS strategy is to construct jets from calorimetric topo-clusters and use jet area subtraction for pileup suppression. In 2016 the experiment relied on a simulation-based energy calibration procedure, whilst for 2017 they also deployed a set of data-driven dijet $\eta$ intercalibration corrections and a procedure for global sequential corrections that, based on the jet longitudinal shape and its associated tracks characteristics, enhance the resolution whilst keeping the average jet energy scale unchanged.

\begin{figure}[htb]
\centering
\includegraphics[height=1.8in]{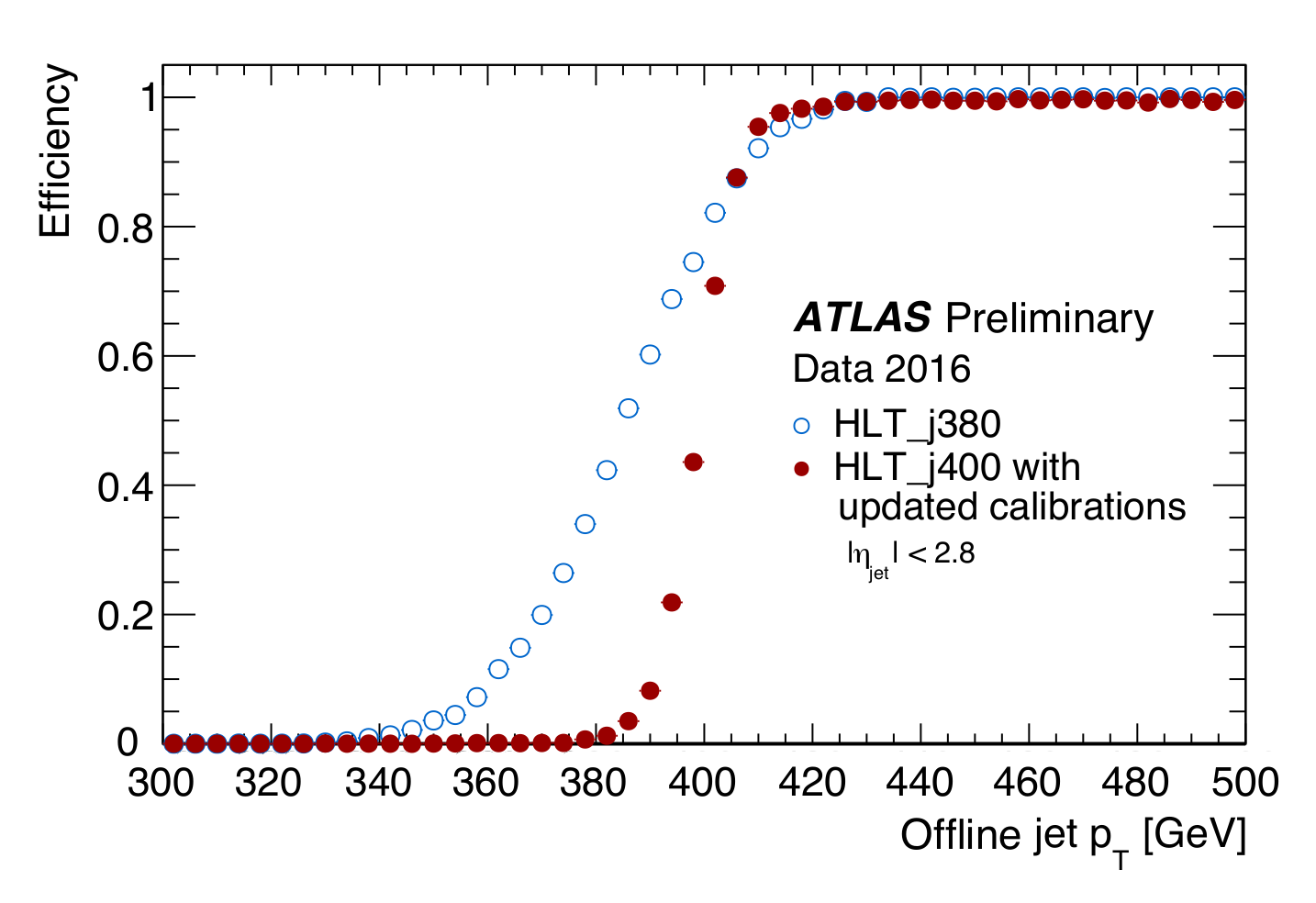}
\includegraphics[height=1.8in]{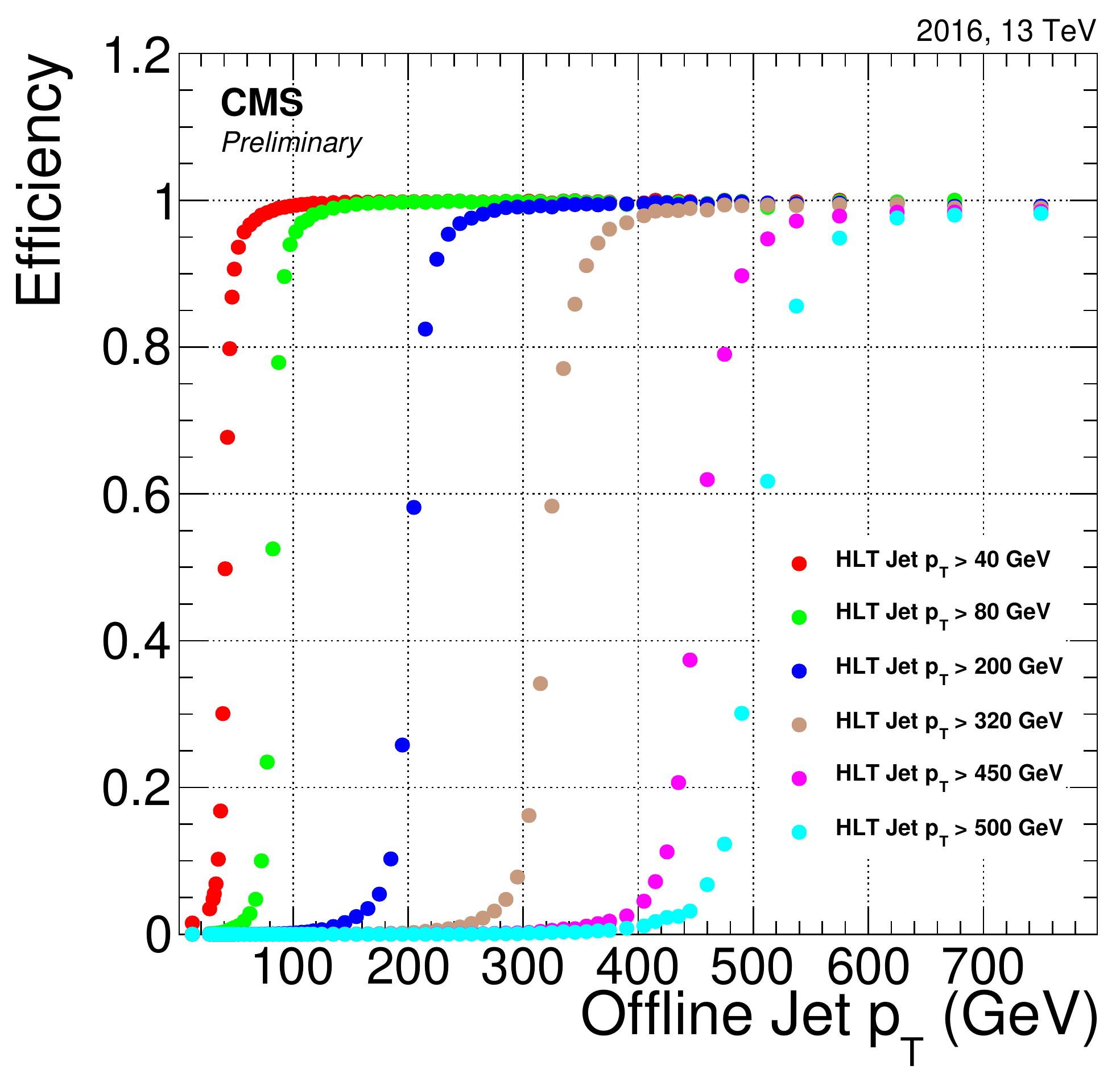}
\setlength{\abovecaptionskip}{7pt}
\caption{Left: ATLAS turn-on curve for the \texttt{HLT\_j380} and \texttt{HLT\_j400} paths, demonstrating the effect of the updated calibrations~\cite{ATLASOverallTRG:2016}.
Right: CMS turn-on curve for different single jet paths, ranging from \PT{} thresholds of 40~GeV to 500~GeV~\cite{CMSHLTRun2:2016}.}
\label{fig:JetTurnOn}
\end{figure}

On the other hand, the CMS strategy relies primarily on the particle-flow (PF) algorithm; in 2016 a large effort was made to align PF between online and offline reconstruction whilst still keeping the former within the timing budget. After a preselection based on calorimetric jets, jets are built from PF candidates. A set of sequential corrections is applied to the jets: a pileup correction, based on the offset event energy density ($\rho$), and a relative correction to make the jet response uniform over $\eta$ and \PT{}. Example performance plots for jet triggers for both experiments can be seen in Fig.~\ref{fig:JetTurnOn}.

\begin{figure}[htb]
\centering
\includegraphics[height=1.65in]{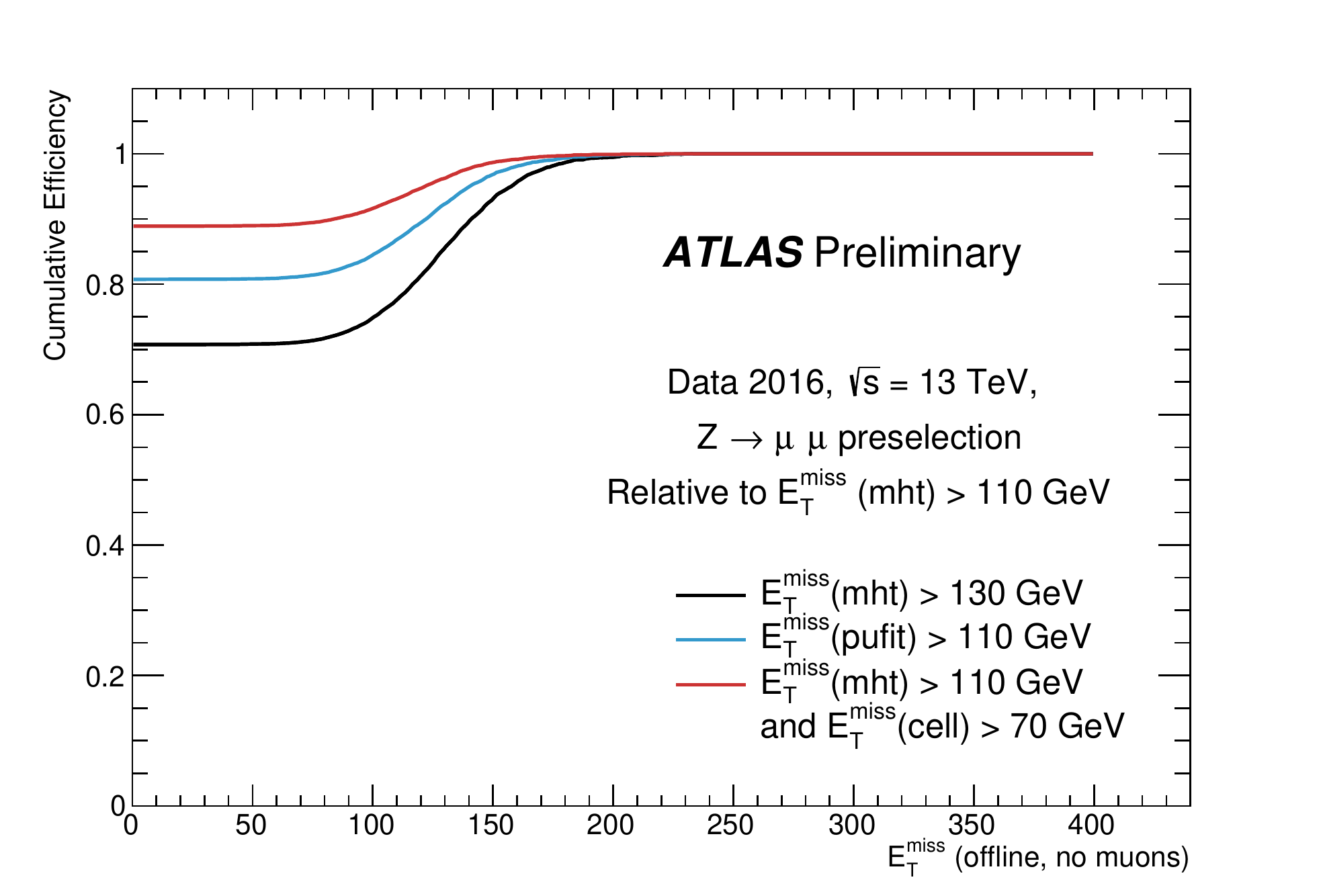}
\includegraphics[height=1.65in]{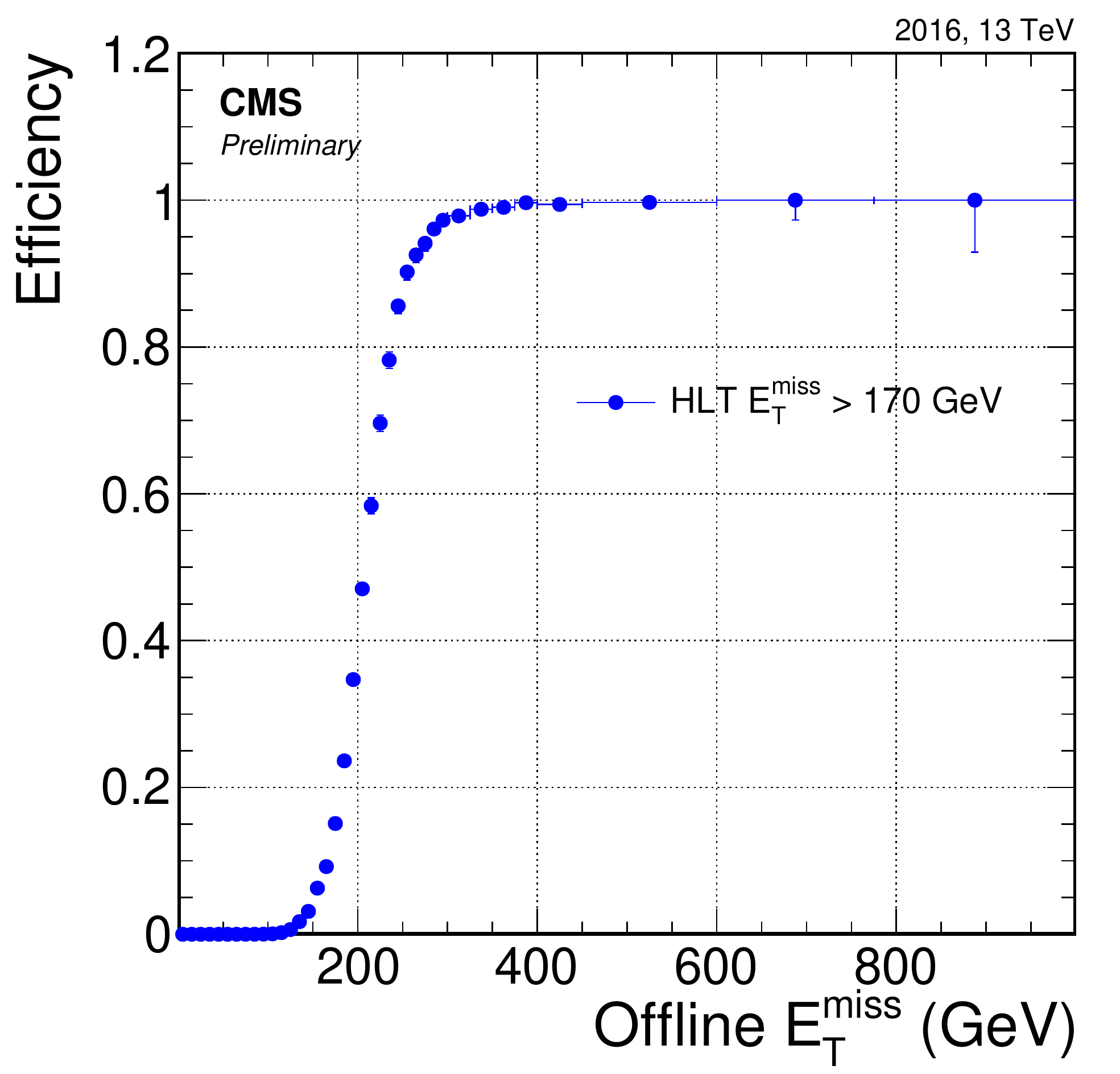}
\setlength{\abovecaptionskip}{7pt}
\caption{Left: ATLAS turn-on curves for different \MET{} algorithms: the \texttt{mht} algorithm reconstructs the \MET{} from jets, whilst the \texttt{pufit} and \texttt{cell} employ additional techniques to minimize the effects of pileup~\cite{ATLASOverallTRG:2016}. Right: CMS turn-on for the \texttt{HLT\_PFMET170} path, seeded by a full set of L1 seeds with thresholds up to 120~GeV~\cite{CMSHLTRun2:2016}. Both offline and online \MET{} are reconstructed with the Particle Flow algorithm.}
\label{fig:METTurnOn}
\end{figure}

Both experiments also deploy missing \ET{} (\MET{}) trigger paths, particularly targeting searches for BSM physics signals like dark matter. ATLAS benefits greatly from their newly unified L2/EF HLT structure, being able to do offline full-detector reconstruction
directly after the L1 trigger decision. They deploy various \MET{} reconstruction methods: whilst for 2015 a topo-cluster based approach was used, in 2016 the default algorithm was based on reconstruction of the \MET{} from jets (\MHT{}). In order to help reduce the trigger rate at high pileup, advanced algorithms that either fit the pileup effect or combine jet-based and cell-based information are being investigated for 2017. CMS again employs a strategy of preselection on calorimetric \MET{} followed by particle-flow \MET{} reconstruction; they also use combined selections on \MET{} and \MHT{} to keep trigger rate under control whilst having high efficiency for events with real momentum imbalance. Finally, both experiments also have trigger paths that select events with high amount of hadronic activity by applying thresholds on the scalar sum of all jets above a given threshold (\HT{}). Example performance plots for missing energy triggers for both experiments can be seen in Fig.~\ref{fig:METTurnOn}.

\section{Other Trigger Paths}

From the basic objects, a multitude of different trigger paths may be constructed targeting specific event topologies. Jet triggers can be enhanced by requesting jets to b-tagged or boosted-tagged; a completely different procedure may be used to select events where an hadronic tau is to be reconstructed instead. \HT{} triggers can be enhanced by optionally selecting a minimum jet multiplicity and/or special tags in the jets. Lepton + jet paths can be used to explore regions of the phase space where requiring the presence of only one object would be prohibitive due to the high thresholds that would be needed.

\vspace*{-4pt}
\section{Conclusions and Outlook}

The LHC Run~2 brought harsher conditions upon the ATLAS and CMS experiments, with 13~TeV centre-of-mass energy and 25~ns bunch spacing for pp collisions. During 2016, the instantaneous luminosity went up to 1.4$\times$10\textsuperscript{34}~cm\textsuperscript{-2}~s\textsuperscript{-1} -- and will increase further in 2017. The trigger systems of both experiments were improved to cope with these conditions whilst maintaining physics performance, both by deploying more powerful hardware as well as using better data reconstruction and selection algorithms. More than 30~fb\textsuperscript{-1} of pp collision data were taken during Run~2 up until now, and improvements are still ongoing. Meanwhile, both experiments are still working on the upgrades for the LHC Run~3 and the High Luminosity LHC.

\Acknowledgements
\supporttext{}

\end{document}